\documentclass[aps,prl,reprint,amsmath,amssymb,superscriptaddress,floatfix]{revtex4-2}

\usepackage{graphicx}
\usepackage{dcolumn}
\usepackage{bm}
\usepackage{float}\usepackage{amssymb}
\usepackage[dvipsnames]{xcolor}
\usepackage{tcolorbox}
\usepackage{amsmath}
\usepackage{tensor}
\usepackage{cases}
\usepackage{hyperref}

\usepackage{braket}

\begin{document}
	
	\title{Thermodynamics of analogue black holes in a non-Hermitian tight-binding model}
	
	\author{D.F. Munoz-Arboleda}
	\email{d.f.munozarboleda@uu.nl}
	\affiliation{Institute for Theoretical Physics, Utrecht University, 3584CC Utrecht, The Netherlands}%
	
	\author{M. St\aa{}lhammar}
	\affiliation{Institute for Theoretical Physics, Utrecht University, 3584CC Utrecht, The Netherlands}
	\affiliation{Department of Physics and Astronomy, Uppsala University, Uppsala, Sweden}

	\author{C. Morais Smith}
	\affiliation{Institute for Theoretical Physics, Utrecht University, 3584CC Utrecht, The Netherlands}%
	
	\date{\today}
	
	\begin{abstract}
	We present a non-Hermitian model with gain/loss and non-reciprocal next-nearest-neighbor hopping that emulates black-hole physics. The model describes a one-dimensional lattice with a smooth connection between regions with distinct hopping parameters. By mapping the system to an effective Schwarzschild metric in the Painlevé-Gullstrand coordinates, we find that the interface is analogue to a black-hole event horizon. We obtain emission rates for particles and antiparticles, the Hawking temperature, the Bekenstein-Hawking entropy, and the mass of the analogue black hole as a function of the interface sharpness and the system parameters. An experimental realization of the theoretical model is proposed, thus opening the way to the detection of elusive black-hole features. 
	
	\end{abstract}
	
	\maketitle

\textit{Introduction.}--- For more than 50 years, one of the most intriguing findings in physics has been the black hole (BH) process \cite{hawking_black_1974}.
Hawking predicted that a quantum field placed in a curved spacetime background that describes a BH will emit thermal radiation \cite{hawking_particle_1975}. 
One of the main difficulties in observing Hawking radiation is that, for the smallest astrophysical BH \cite{chandrasekhar_maximum_1931, oppenheimer_continued_1939, Abbott_long}, the Hawking radiation is extremely weak, making it virtually impossible to observe \cite{carr_black_1974, carr_observational_2024, barcelo_analogue_2011}. 
One way to avoid this difficulty is by looking at primordial BHs (BHs created at the early stages of the universe). 
However, up to now, there is no record of this type of BH. 

In 1981, W. G. Unruh introduced the concept of analogue gravity by showing that sound waves in a moving fluid could be described by an effective metric analogue to that of a BH, including an acoustic horizon and associated Hawking-like radiation \cite{unruh_experimental_1981}.
This insight established the foundation for a broad interdisciplinary research program that demonstrated how controllable laboratory systems could emulate aspects of quantum field theory in curved spacetime. 
Subsequent theoretical developments extended this framework to a variety of media, identifying analogues of phenomena such as superradiance, cosmological particle production, and horizon backreaction in systems governed by effective metrics derived from fluid dynamics, optics, and condensed matter \cite{BarceloLiberatiVisser-LRR2011, weinfurtner_measurement_2011, subramanyan_physics_2021, patrick_backreaction_2021}.
Experimental efforts confirmed these theoretical predictions, including the measurement of stimulated Hawking radiation in surface water waves under controlled flow conditions \cite{weinfurtner_measurement_2011, patrick_backreaction_2021}, as well as the generation of horizon-like effects using optical fibers and slow-light systems. 
In particular, the observation of spontaneous and entangled phonon emission in Bose–Einstein condensates provided strong evidence for quantum Hawking radiation in an analogue setting \cite{steinhauer_observation_2016}.

While these developments have been grounded in Hermitian physics, their extension to non-Hermitian (nH) systems has opened new avenues for understanding spacetime analogues and causal structures in dissipative or driven settings. 
Most traditional analogue BH implementations rely on real-valued wave equations and conserve probability, which restricts the range of phenomena that can be modeled. 
However, many realistic experimental platforms, including those involving gain/loss and amplification, operate in nH regimes.
Recently, it was shown that nH Dirac-like systems with parity-time ($\mathcal{PT}$) symmetry can exhibit structures analogue to light cones in general relativity, with a well-defined effective line element that mimics causal flow near horizons \cite{longhi2020non,Stalhammar-NJP2023}. 
This approach establishes a formal bridge between nH band structures and spacetime geometry, allowing phenomena such as horizon-induced tunneling or spectral flow to be recast in terms of complexified metrics. 
These results suggest that the inclusion of non-Hermiticity is not merely a perturbative feature but can play a central role in defining new classes of analogue gravitational systems.

The Painlevé-Gullstrand formulation of the Schwarzschild geometry is particularly well-suited for analogue gravity models, as it allows for a clear description of radial flow and horizon structure without coordinate singularities \cite{Painleve-1921, Volovik-JETPL1999, Stalhammar-NJP2023}. 
In this coordinate system, the spacetime is foliated by spatial hypersurfaces that are flat but threaded by a velocity field representing the infall of a freely falling observer.
This form of the metric bears a striking resemblance to the energy dispersion of nH lattice models, where the non-reciprocal hopping induces a directional energy flow and an effective group velocity that can be interpreted as a lattice analogue of gravitational infall. 
These coordinates are particularly important for describing Hawking and Unruh radiation, as they allow for a regular, observer-dependent formulation of horizon-crossing processes.

The Parikh–Wilczek approach provides an alternative perspective on Hawking radiation by modeling it as a quantum tunneling process across the BH horizon, rather than as a phenomenon strictly related to global spacetime features \cite{ParikhWilczek-PRL2000}. 
In this semiclassical formulation, particles escape from the horizon through classically forbidden regions, and the resulting emission spectrum is nearly thermal, with corrections due to energy conservation and backreaction.
This framework refines the original derivation by accounting for the self-interaction of the tunneling particle, and has become central in discussions of BH thermodynamics and quantum corrections to evaporation. 
In analogue gravity systems, this formalism offers a powerful tool to interpret emission processes as effective tunneling events, especially in contexts where the underlying dispersion relations are modified or complexified, as occurs in dispersive media or nH platforms.

In this work, we explore these connections by studying a nH tight-binding (TB) model that includes on-site gain/loss, as well as non-reciprocal next-nearest-neighbor (NNN) hopping.
We show that this model exhibits a regime in which the NNN hopping across a spatial interface emulates the behavior of frequency modes near a BH horizon in Painlev\'e-Gullstrand coordinates. 
By analyzing the complex-valued dispersion relation near the interface, we demonstrate the emergence of an effective tunneling process, with an emission rate that is similar to that derived by Parikh-Wilczek \cite{ParikhWilczek-PRL2000}. 
This allows us to define a Hawking-like temperature and an associated Bekenstein-Hawking entropy from the lattice spectrum, revealing how thermodynamic features of BH evaporation can emerge from nH band structures.

This correspondence opens a novel route to exploring gravitational phenomena in controlled, table-top settings using engineered nH materials.
While analogue gravity models have been constructed in various fluid systems, Bose-Einstein condensates, and optical media \cite{rosenberg_optical_2020, drori_observation_2019}, our approach highlights the potential of nH-TB lattices to capture not only kinematic aspects of BH analogues but also \textit{thermodynamical} features. 
In particular, the incorporation of gain/loss enables a direct encoding of dissipative effects and information flow, which are essential to any quantum theory of gravity.

From a condensed-matter perspective, our findings illustrate how nH Hamiltonians can simulate relativistic geometries and thermodynamical processes beyond equilibrium.
The emergent geometry in our model is not imposed externally but arises from the interplay of microscopic lattice parameters, suggesting new ways to engineer artificial spacetimes with tunable properties. 
This may pave the way for future investigations into analogue cosmology, topological gravity, and the holographic duality in nH settings.

\begin{figure}[bht]
	\centering
	\includegraphics[scale=1.5]{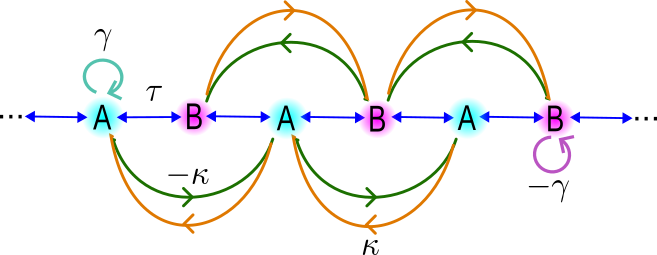}
	\caption{Sketch of the nH-TB model. The blue lines describe the hopping parameter $\tau$, the cyan (magenta) lines represent the gain(loss) potential $\gamma$ ($-\gamma$), and the green (orange) lines describe the non-reciprocal NNN hopping $-\kappa$ ($\kappa$).}
	\label{extended_nHSSH}
\end{figure}

\begin{figure}[bht]
	\centering
	\includegraphics[width=8.8cm, height=2.5cm]{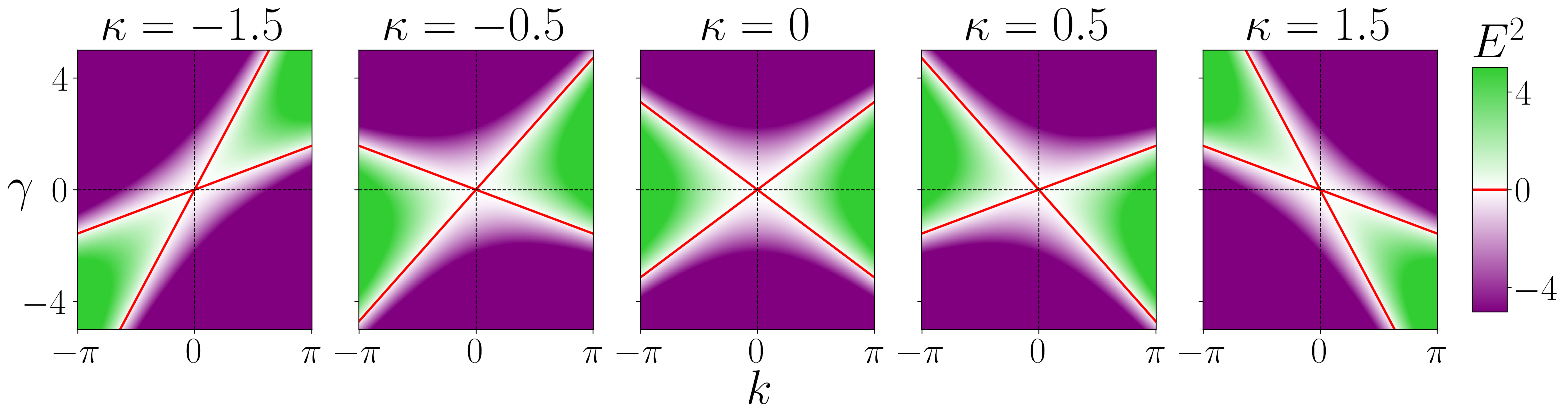}
	\caption{Behavior of the energy square (see color code) as a function of $\gamma$ and $k$ for different values of $\kappa$. For simplicity, $\tau=1$. Red lines indicate the exceptional points}
	\label{gammakenergy}
\end{figure}

\textit{nH-TB model.}--- 
Non-Hermitian extensions of one-dimensional (1D) chains have garnered significant attention, particularly under $\mathcal{PT}$-symmetric conditions \cite{Bender-PRL1998}, where balanced gain and loss potentials lead to novel topological phenomena \cite{BERGHOLTZBUDICH-RMP2021,yang_homotopy_2024, SlootmanCherifi2024}. 
Here, we consider a $\mathcal{PT}$-symmetric nH-TB Hamiltonian $H$ with $N$ unit cells and two sublattices, A (cyan) and B (magenta) (see Fig.~\ref{extended_nHSSH}) 
\begin{align}
\displaystyle H&= \sum_{j=1}^{N}\left[ \tau\left(b_j^{\dagger}a_j +  a_{j+1}^{\dagger}b_j+ \textnormal{h.c.}\right)+i\gamma \left( a_j^{\dagger}a_j-b_j^{\dagger}b_j \right) \right.\nonumber \\
    &- \left. \frac{\kappa}{2}\left(a_{j+1}^{\dagger}a_j-a_{j}^{\dagger}a_{j+1}+b_j^{\dagger}b_{j+1}-b_{j+1}^{\dagger}b_j\right) \right], 
	\label{egl2}    
\end{align}
that incorporates a hopping parameter $\tau$, with $a_{j}^{\dagger}$ and $b_{j}^{\dagger}$ ($a_{j}$ and $b_{j}$) the creation (annihilation) operators on sites A and B of the $j_{\textnormal{th}}$ unit cell, respectively, a balanced gain/loss [$\gamma/(-\gamma)$] in the A and B sublattice, and a non-reciprocal NNN hopping parameter $\kappa$. 
Although there is no difference between the intracell and intercell hopping, it is possible to differentiate the sublattices A and B through the gain and loss potentials and the non-reciprocal NNN hopping.
The lattice constant is put to unity. 
Using periodic boundary conditions (PBC), we derive the Bloch Hamiltonian and expand $\sin(k)$ and $\cos(k)$ around $\pi$ \footnote{If we expand around zero, we would obtain an extra term $2\tau$ and the calculations lead a Dirac-like operator that gives an exceptional cone with a shift in the momentum and does not change the physics of the system} to linear order
\begin{equation}
	h(\tilde{k})=\begin{pmatrix}
		i(\gamma+\kappa \tilde{k})&+i\tau \tilde{k}\\
		-i\tau\tilde{k} & -i(\gamma+\kappa \tilde{k}) 
	\end{pmatrix},
	\label{egl5}
\end{equation}
where $\tilde{k}=(k-\pi)$. 
Here, $h(\tilde{k})$ is $\mathcal{PT}$-symmetric [$h(\tilde{k})=\sigma^xh^*(\tilde{k})\sigma^x$] \cite{MontagKunst-JMP2024}. 
The eigenvalues are 
\begin{equation}
	\epsilon_{\pm}(\tilde{k})=\pm\sqrt{\tau^2\tilde{k}^2-(\gamma+\kappa \tilde{k})^2},
	\label{egl6}    
\end{equation}
and the system hosts exceptional points when $\gamma=-\kappa \tilde{k} \pm |\tau\tilde{k}|$, which form a cone with a tilting profile controlled by $\kappa$, see Fig.~\ref{gammakenergy}. 
A similar tilting profile can be found in the surroundings of a Schwarzschild BH event horizon in Painlev\'e-Gullstrand coordinates by tuning the free falling velocity~\cite{Painleve-1921, Hamilton-AJP2008}. 
Hence, a relation between nH $\mathcal{PT}$-symmetric models and BHs becomes evident~\cite{Stalhammar-NJP2023}. 
For $\kappa=0$, we see a cone of exceptional points, which is tilted by the parameter $\kappa$. 
If $|\kappa|>1$, the cone is over-tilted (similar to a Lifshitz transition in Weyl materials)~\cite{volovik_topological_2017, wu_topological_2023}.

\textit{Analogue spacetime.}--- To relate the exceptional cone (see Fig.~\ref{gammakenergy} red lines) with the light cone of a Schwarzschild BH, we follow the procedure outlined in Ref.~\cite{Stalhammar-NJP2023}. 
Noticing that $\gamma$ takes the form of the eigenvalues of a Dirac operator, we promote it to an operator,
\begin{equation}
	\hat{\gamma}=-\kappa \tilde{k} \sigma^0+|\tau\tilde{k}|\sigma^x \to \hat{\gamma}=e\indices{^\mu_\alpha} \tilde{k}_\mu\sigma^\alpha,
	\label{m2}
\end{equation}
where $e\indices{^\mu_\alpha}$ are \textit{vielbeins} or tetrad fields~\cite{Carroll-2019, Ortin-2004, VOLOVIK-NPB2014, Horava-PRL2005}, $\mu,\alpha\in\{0,1\}$, $\sigma^{0}=\mathbf{1}$, and $\sigma^1$ is a Pauli matrix. 
The \textit{vielbeins} give rise to a metric of the form $g^{\mu\nu}=e\indices{^\mu_\alpha} e\indices{^\nu_\beta} \eta^{\alpha\beta}$ where $\eta^{\alpha\beta}=\textnormal{diag}(-1,1)$. 
Thus, a line element of the spacetime is obtained as $ds^2=g_{\mu\nu}dx^{\mu}dx^{\nu}$. 
For the nH-TB model defined above, the \textit{vielbeins} are $e\indices{^0_0}=1$, $e\indices{^1_1}=\tau$, and $e\indices{^1_0}=-\kappa$. 
Without losing generality, we fix $\tau=1$. Then, the line element can be written as
\begin{equation}
	ds^2 = -\left(1 - \kappa^2 \right)(dt)^2 + 2 \kappa dt dx + (dx)^2,
	\label{m4}
\end{equation}
which is similar to the Schwarzschild metric in the Painlev\'e-Gullstrand coordinates
\begin{equation}
	ds^2 = -\left(1 - \frac{2M_o}{r} \right)(dt_r)^2 + 2 \sqrt{\frac{2M_o}{r}} dt_r dr,
	\label{m5}
\end{equation}
with $M_o$ denoting the mass at the event horizon for a freely falling observer \cite{ParikhWilczek-PRL2000,Painleve-1921,Lemaitre-1933,Vanzo-CQG2011,Volovik-JETPL1999}. 
In Eq.~\eqref{m4}, if $\kappa$ is positive (negative) we obtain the analogue of a black (white) hole.  

\textit{Analogue event horizons through coupled chains.}--- 
Since the nH-TB model described above can be related to the metric of a Schwarzschild BH in the Painlev\'e-Gullstrand coordinates, we now construct a model that allows us to mimic the corresponding event horizon. 
To do so, we couple two nH-TB chains with on-site gain/loss $\gamma$, a hopping parameter $\tau=1$, and NNN hopping $\kappa_1$ and $\kappa_2$, respectively. 
The two chains are connected by smooth interfaces with a kink (anti-kink) at position $r_1$ ($r_2$), and the NNN hopping across the interfaces is $\pm\kappa_3$.  

\begin{figure*}[tbh]
	\centering
	\includegraphics[scale=1.35]{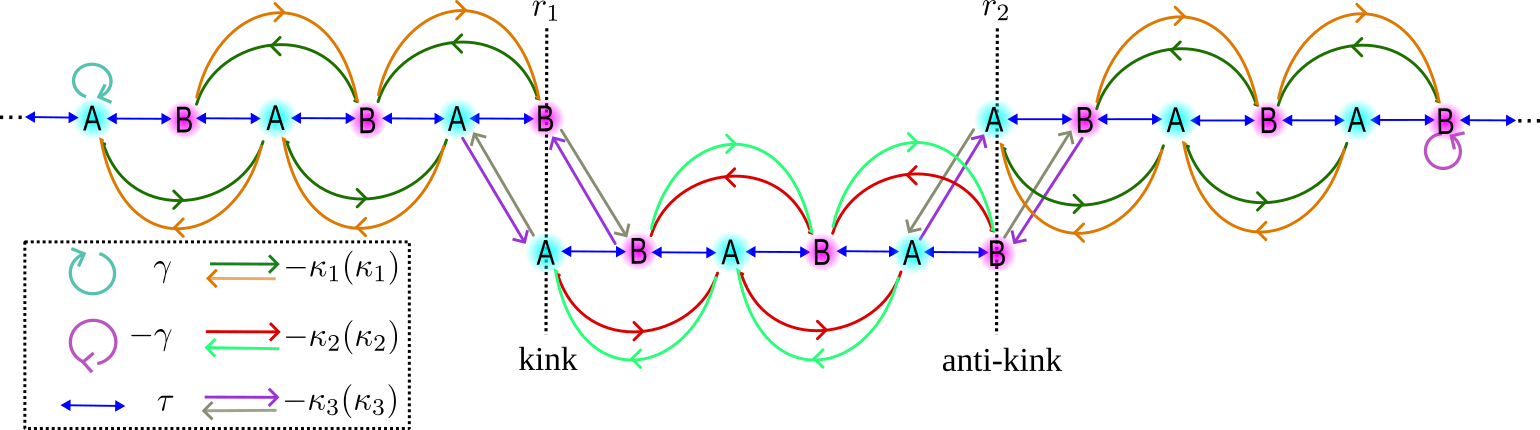}
	\caption{Sketch of the two coupled gain/loss nH-TB chains with non-reciprocal NNN hopping. The blue line represents the nearest neighbor hopping parameter $\tau$. The green (orange) line represents the NNN hopping parameter $\kappa_1$ ($\kappa_2$) for the first chain (second chain). The purple (gray) line represents the NNN hopping parameter ($\kappa_3$) between A (B) sites through the interface between the first and second chain, mimicking the horizon of the BH.}
	\label{bhlc}
\end{figure*}

The setup is illustrated in Fig.~\ref{bhlc}, where we assume PBC. 
The first chain represents the inside of the BH, and the second chain represents the outside. 
The kink (anti-kink) represents the transition point in real space from the first to the second (second to the first) chain, which is analogue to an event horizon. 
The Bloch Hamiltonian describing this system reads
\begin{equation}
	h(k)=\begin{pmatrix}
		i\left[\gamma-f(r)\sin(k)\right]&1 + e^{-i k}\\
		1 + e^{i k}& -i\left[\gamma-f(r)\sin(k)\right] 
	\end{pmatrix},
	\label{bhlc1}
\end{equation}
where $f(r)$ is assumed to have the form
\begin{equation}
	f(r)=\kappa_1+\frac{\kappa_1-\kappa_2}{2}\left[ \textnormal{th}\frac{(r_1-r)}{l}
	+\textnormal{th}\frac{(r-r_2)}{l}\right],
	\label{bhlc4}
\end{equation}
and thus readily include the NNN hopping of each individual chain: $f(r)\sim \kappa_1$ for $r<r_1$, $r>r_2$, and $f(r)\sim\kappa_2$ for $r_1<r<r_2$), and $f(r_1)=f(r_2)=(\kappa_1+\kappa_2)/2=\kappa_3$. 
In Eq.~\eqref{bhlc4}, $l$ is the width of the domain wall. 
Performing the same expansion as in Eq.~\eqref{egl5}, we find $h(\tilde{k})=-\tilde{k}\sigma^y+i\left[\gamma+f(r)\tilde{k}\right]\sigma^z$.
The corresponding cone of exceptional points is described by $\gamma=-f(r)\tilde{k}\pm|\tilde{k}|$.

However, the explicit spatial dependence of the function $f(r)$ makes the commutation relation $[\tilde{k},h(\tilde{k})]$ non-vanishing, since
\begin{equation}
	\left[\tilde{k},h(\tilde{k})\right]=\tilde{k}\frac{\partial f(r)}{\partial r}.
	\label{sler1}
\end{equation}
We have neglected these contributions when solving the eigenvalue equation. 
Close to the analogue horizon $r\rightarrow r_1$ in a sufficiently large chain, i.e.  $r_1\ll r_2$, this amounts to assuming
\begin{eqnarray}
	\displaystyle \left|\frac{\kappa_2-\kappa_1}{2l}\tilde{k}\right|\ll|\gamma|, \quad \left|\frac{\kappa_2-\kappa_1}{2l}\right|\ll|\tau|,
    \label{sler2}
\end{eqnarray}
which define the parameter ranges in which the lattice model mimics the interior and exterior of a BH, as well as the corresponding event horizon. 

Following the same procedure as in the previous section, the metric for this particular model is
\begin{equation}
	ds^2=-\left[1-f^2(r)\right]dt^2+2f(r)drdt+dr^2.
	\label{bhlc5}
\end{equation}
We now want to find where Eq.~\eqref{bhlc5} describes a Schwarzschild-like BH. 
This can be done by expanding Eq.~\eqref{m5} and Eq.~\eqref{bhlc5} to linear order in $(r-2M_o)$ and $(r-r_1)$, respectively, and equating order by order to ensure identical behaviors at and close to the horizon,
\begin{equation}
	\frac{\kappa_1+\kappa_2}{2}-\frac{\kappa_1-\kappa_2}{2l}(r-r_1)=1-\frac{1}{4M_o}(r-2M_o).
	\label{bhlc7}
\end{equation}
Equation~\eqref{bhlc7} defines the horizon in terms of the model parameters, namely
\begin{equation}
    \kappa_3=\frac{\kappa_1+\kappa_2}{2}=1,\  
     r_1=2M_o,\ \textnormal{and}\  
    \frac{\kappa_1-\kappa_2}{2l}=\frac{1}{2r_1},    
    \label{bhlc8}
\end{equation}
which bounds the NNN hopping parameters to $\kappa_1>1$, $\kappa_2<1$, and $\kappa_3=1$.

\textit{Semiclassical limit and emission rate.}---
Having established the relation between our nH-TB model and the Schwarzschild BH, we now turn to the corresponding BH evaporation. 
This is done by studying the tunneling process close to the horizon using the Parikh-Wilczek method \cite{ParikhWilczek-PRL2000}, which takes two processes contributing to the spontaneous emission of massless particles into account. 
One is obtained from a particle tunneling outwards the interior of the BH, and the other one comes from an antiparticle tunneling into the BH. 
In Fig.~\ref{gammakenergy}, we depict with red lines the region in momentum space that corresponds to the exceptional cone, which is analogue to the light-like shell of energy seen by a radially falling observer. 
The momentum of a quantum massless particle or antiparticle follows the radial falling trajectories (radial null geodesics) \cite{birrell1982QFTCS,wald1994QFTCST}, hence $\dot{r}^{\pm} \rightarrow p_{r}^{\pm}$. 
In our lattice model, $\tilde{k}^{\pm}$ is the analogue of the conjugate momentum $p_{r}^{\pm}$. 
On the exceptional cone, we obtain
\begin{equation}
    \tilde{k}^{\pm}_r=\frac{\gamma f(r)}{1-f^2(r)}\pm \sqrt{\frac{\gamma^2}{\left[1-f^2(r)\right]^2}}.
\end{equation}
Now, we need to identify which solution is related to the analogue of the particle and the antiparticle channel. 
Usually, the method for determining the particle and antiparticle channels comes from the behavior of $\tilde{k}^{\pm}(r)$ when approaching the asymptotic infinity. 
In our analogue model the asymptotic infinity lies in $r_1<r<r_2$; it is bound because of the PBC. 
For a sufficeintly large second chain, there exists an $r_{\infty}$ such that $r_1\ll r_{\infty}\ll r_2$, and consequently $f(r_{\infty})\sim \kappa_2$. 
Then, we get
\begin{equation}
    \tilde{k}_r^{\pm}(r_\infty) = \frac{\gamma \kappa_2}{1-\kappa_2^2}\pm\Biggl\{\frac{\gamma^2}{\left[1-\kappa_2^2\right]^2}\Biggr\}^{1/2}.
    \label{sler4}
\end{equation}
From the conditions given in Eq.~\eqref{bhlc8}, we know that $0<\kappa_2<1$ and obtain $\tilde{k}_r^{+}\geq0$ and $\tilde{k}_r^{-}\leq0$. 
These equations describe the analogue momentum of a particle ($\tilde{k}_r^{+}$) and an antiparticle ($\tilde{k}_r^{-}$) for any real value of $\gamma$. 
The classically forbidden process is obtained by computing the imaginary part of the action $\textnormal{Im}(S^{\pm})=\textnormal{Im}\left(\int dr \tilde{k}_r^{\pm}\right)=I^{\pm}$ \cite{ParikhWilczek-PRL2000,Stalhammar-NJP2023}. 
These solutions correspond to the particle ($+$) and antiparticle ($-$) channels involved in the tunneling process; the particle channel describes a positive-frequency ($\omega>0)$ particle tunneling out from the BH, while the antiparticle channel describes a negative-frequency ($\omega'<0$) antiparticle tunneling into the BH.  
As explained above, this implies that (anti)particles radiate out from (into) the BH interior, causing the BH mass to decrease from $M_o$ to $M_o - \omega$ ($M_o+\omega')$, which in turn leads to a shrinking of the horizon. 
Taking the change in mass into account, the integration limits around the horizon are set as $r_{\textnormal{in}} = r_1$ and $r_{\textnormal{out}} = r_1 - 2\omega$ ($r_{\textnormal{out}} = r_1 + 2\omega'$), where $r_1=2M_o$ is obtained from Eq.~\eqref{bhlc7}. 
Inserting $\tilde{k}^{\pm}_r$ into $\textnormal{Im}(S^{\pm})$, we obtain $I^{\pm}$. 
By solving the integral for the particle and antiparticle channel, and recalling that both channels contribute to the probability of the tunneling process \cite{ParikhWilczek-PRL2000}, we obtain the semiclassical emission rate \cite{ParikhWilczek-PRL2000,Stalhammar-NJP2023}
\begin{equation}
    \Gamma\propto (\mathcal{A}^+ + \mathcal{A}^-)^2= e^{-8\pi M_o|\gamma|}=e^{\Delta S_{\textnormal{B-H}}},
    \label{sler12}
\end{equation}
where $\mathcal{A}^{\pm}=\exp(-I^{\pm})$. 
Detailed calculations are provided in a Supplemental Material~\footnote{See Supplemental Material for a detailed calculation of the imaginary part of the action}. The above equation shows the relation between the exponential part of the semiclassical emission rate and the change in the Bekenstein-Hawking entropy ($\Delta S_{\textnormal{B-H}}$) \cite{KeskivakkuriNPB1997,MASSARNPB2000,ParikhWilczek-PRL2000, bekenstein_black_1972, bekenstein_black_1973, hawking_particle_1975}, which leads to
\begin{equation}
    \Delta S_{\textnormal{B-H}}=-8\pi M_o|\gamma|=-4\pi |\gamma|\frac{l}{\kappa_1-\kappa_2}.
    \label{sler13}
\end{equation}

\textit{Thermodynamics of the analogue BH.}--- 
To complement the calculation of $\Delta S_{\textnormal{B-H}}$, we compute the difference in the thermal entropy $\Delta S$ from thermodynamic principles $dM=T_\textnormal{H}dS$, where $T_\textnormal{H}=1/8\pi M$ is the Hawking temperature of a Schwarzschild BH. 
Comparing Eqs.~\eqref{m5} and \eqref{bhlc5} term by term, we can obtain a relation between the mass $M$ of the BH and the function $f(r)$ that describes the behavior of the NNN hopping through the chains. 
Notice that we are interested in a variable mass $M$ and not only in the mass at the event horizon $M_o$. 
The thermal entropy is written as
\begin{equation}
    \Delta S=\int\limits_{M_o}^{M_o-\omega}8\pi M dM = -\frac{3\pi}{2}\left(\frac{l}{\kappa_1-\kappa_2}\omega+\omega^2\right).
    \label{tabh2}
\end{equation}
Equating $\Delta S_{\textnormal{B-H}}=\Delta S$, we get $-8\pi M|\gamma|=-3\pi(2M\omega+\omega^2)/2$. 
In terms of the lattice parameters, the frequency reads
\begin{equation}
    \omega^{\pm}=-\frac{l}{2(\kappa_1-\kappa_2)}\pm\sqrt{\frac{l^2}{4(\kappa_1-\kappa_2)^2}+\frac{8l|\gamma|}{3(\kappa_1-\kappa_2)}}.
    \label{tbah6}
\end{equation}
Recalling Eq.~\eqref{sler2}, $(\kappa_1-\kappa_2)/2l \ll 1$, the leading order contribution to the frequency becomes $\omega^+=8|\gamma|/3$. 
This remarkably simple relation indicates that the frequency of the emitted particles responsible for the BH evaporation, and hence the semiclassical emission rate, is controlled exclusively by the gain/loss parameter.
Practically, this means that the nature of different BH evaporation processes can be scanned through in our model by simply tuning the on-site potential.

\textit{Discussion.}--- 
We introduced a nH-TB model with $\mathcal{PT}$-symmetric gain/loss and non-reciprocal NNN hopping, which gives rise to exceptional cones with tunable tilt. 
By expanding the Bloch Hamiltonian near the Brillouin zone edge, we revealed Dirac-like operators that mimic light cones in curved spacetime. 
Mapping the cone tilt to Painlev\'e-Gullstrand coordinates, we identified a correspondence with Schwarzschild black holes. 
This analogy enables us to design a tabletop setup with two smoothly connected chains, emulating a black hole horizon with interior, exterior, and interface regions.

To examine tunneling across the analogue horizon, we employed a semiclassical approach inspired by the Parikh-Wilczek method \cite{ParikhWilczek-PRL2000}, deriving an emission rate governed by the change in B-H entropy. 
Additionally, by analyzing the thermodynamic response of the system, we obtained a compact expression linking the emission frequency to the on-site gain/loss parameter, $\omega^+ = 8|\gamma|/3$. The quadratic contribution in our entropy balance, $-8\pi M|\gamma|=-3\pi(2M\omega+\omega^2)/2$ is a consequence of the energy conservation in the tunneling process, in direct analogy with the Parikh--Wilczek correction to strict thermality.

Our results provide a concrete example of how nH topological models can simulate aspects of BH physics, including horizon dynamics and Hawking-like radiation. 
In a tabletop realization, the analogue “particle/anti-particle” channels correspond to the correlated creation of two excitations in distinct mode sectors generated by horizon-induced mode conversion (e.g., opposite group velocities or opposite branches of the effective dispersion), rather than to literal matter-antimatter emission.
Operationally, these channels are therefore most naturally identified through correlations between signals measured on the exterior and interior sides of the interface or between two frequency bands, accessible via transmission and reflection spectra, time-resolved wavepacket scattering, and two-point correlation measurements.
Moreover, since the horizon location is set by $r_1=2M_o$ and the emitted energy lowers the effective mass to $M_o-\omega$ with $\omega=8|\gamma|/3$, the corresponding time-dependent motion of the interface can be emulated experimentally by tracking the evolution of a Gaussian wavefunction or a two-point spatial correlation function across the horizon.

The implementation of our model in an experimental setup is of great interest because it can yield insight into the nature of BH evaporation, the emission rate due to tunneling processes near the event horizon, and the thermodynamics of such systems. 
Platforms capable of realizing gain/loss potentials and non-reciprocal contributions simultaneously include photonic crystals \cite{Reisenbauer-Nature2024}, topoelectric circuits \cite{lee_topolectrical_2018, PhysRevB.109.115407}, and robotic metamaterials \cite{BrandenbourgerCorentin-Nature2019}, to cite just a few. 
By measuring the local density of states in the model described here, a relation between the probability of the emission rate of particles and antiparticles, the Hawking temperature, and the thermal entropy can be obtained.

\textit{Acknowledgments.}--- The authors thanks R. Arouca for the valuable discussions.
DFM-A acknowledges funding from the Colombian Ministry of Science (Minciencias) and CMS acknowledges funding from QuMat, a program of the Netherlands organization for Scientific Research (NWO) that is funded by the Dutch Ministry of Education, Culture and Science.  
M.S. is supported by the Swedish Research Council (VR) under Grant No. 2024.00272. 

\bibliography{ThnHTBABs}
\clearpage
\appendix

\onecolumngrid

\begin{center}
\textbf{\large Supplemental Material: Thermodynamics of analogue black holes in a non-Hermitian tight-binding model}
\vskip 0.5em
D.F. Munoz-Arboleda, M. Stålhammar, and C. Morais Smith\\
\textit{Institute for Theoretical Physics, Utrecht University, 3584CC Utrecht, The Netherlands}\\
\textit{Department of Physics and Astronomy, Uppsala University, Uppsala, Sweden (M.S.)}
\end{center}

\vspace{1em}
\setcounter{equation}{0}
\setcounter{figure}{0}
\setcounter{table}{0}
\renewcommand{\theequation}{S\arabic{equation}}
\renewcommand{\thefigure}{S\arabic{figure}}
\renewcommand{\thetable}{S\arabic{table}}

\section{I. Review of Parikh--Wilczek tunneling formalism}
\label{sec:PWsummary}

This section summarizes the Parikh-Wilczek derivation of Hawking radiation as a semiclassical tunneling process \cite{ParikhWilczek-PRL2000,KrausWilczek-NPB1995}.
We work in units $G=c=\hbar=k_B=1$ and restrict to s-wave, massless emission.

A convenient coordinate system for the tunneling calculation is the Painlev\'e-Gullstrand form of the Schwarzschild geometry, regular at the horizon,
\begin{equation}
ds^2
= -\left(1-\frac{2M_o}{r}\right)dt^2
+2\sqrt{\frac{2M_o}{r}}\,dt\,dr
+dr^2
+r^2 d\Omega^2,
\label{PW_PG_metric}
\end{equation}
where $M_o$ is the black-hole mass and $d\Omega^2$ the line element on the unit sphere.
For radial null trajectories ($ds^2=0$, $d\Omega^2=0$), one finds
\begin{equation}
\dot r \equiv \frac{dr}{dt}
= \pm 1 - \sqrt{\frac{2M_o}{r}},
\label{PW_null_geodesics}
\end{equation}
where the upper sign corresponds to outgoing modes (which stall at the horizon). 

In the Parikh-Wilczek picture, emission is treated as a classically forbidden trajectory across the horizon.
The tunneling probability is controlled by the WKB factor
\begin{equation}
\Gamma(\omega)\propto \exp\!\Bigl[-2\,\mathrm{Im}\,S(\omega)\Bigr],
\label{PW_rate_def}
\end{equation}
where $S$ is the classical action for an outgoing shell of energy $\omega$.
Writing the action in Hamilton--Jacobi form,
\begin{equation}
\mathrm{Im}\,S
=\mathrm{Im}\int_{r_{\rm in}}^{r_{\rm out}} p_r\,dr
=\mathrm{Im}\int_{r_{\rm in}}^{r_{\rm out}}
\int_{0}^{p_r} dp'_r\,dr,
\label{PW_action_pr}
\end{equation}
and using Hamilton's equation at fixed $r$,
\begin{equation}
\dot r=\left.\frac{dH}{dp_r}\right|_{r},
\label{PW_Hamilton_eq}
\end{equation}
one converts the momentum integral into an energy integral,
\begin{equation}
\mathrm{Im}\,S
=\mathrm{Im}\int_{r_{\rm in}}^{r_{\rm out}}
\int_{H_{\rm in}}^{H_{\rm out}}
\frac{dH}{\dot r}\,dr.
\label{PW_action_energy}
\end{equation}

Energy conservation is implemented by allowing the background to change during the emission:
as a shell of instantaneous energy $\omega'$ tunnels out, the mass parameter becomes
\begin{equation}
M\ \longrightarrow\ M-\omega'.
\label{PW_backreaction}
\end{equation}
Therefore, the outgoing radial null equation along the tunneling path is
\begin{equation}
\dot r = 1-\sqrt{\frac{2(M-\omega')}{r}}.
\label{PW_rdot_backreaction}
\end{equation}
The integrand in Eq.~\eqref{PW_action_energy} has a simple pole at the (instantaneous) horizon
$r_H(\omega')=2(M-\omega')$.
Deforming the $r$-contour around this pole with the standard Feynman prescription yields the imaginary part.
Carrying out the $(r,\omega')$ integrals gives the Parikh-Wilczek result
\begin{equation}
\mathrm{Im}\,S
=4\pi\,\omega\!\left(M-\frac{\omega}{2}\right),
\label{PW_ImS_result}
\end{equation}
so that
\begin{equation}
\Gamma(\omega)\propto
\exp\!\left[-8\pi\,\omega\!\left(M-\frac{\omega}{2}\right)\right].
\label{PW_rate_result}
\end{equation}
The leading term ($\omega\ll M$) reproduces a thermal Boltzmann factor,
\begin{equation}
\Gamma(\omega)\propto e^{-\omega/T_{\rm H}},
\qquad
T_{\rm H}=\frac{1}{8\pi M},
\label{PW_thermal_limit}
\end{equation}
while the $\mathcal O(\omega^2)$ correction encodes the deviation from exact thermality due to energy conservation.

A compact and widely used form of the result is obtained by rewriting the exponent as an entropy difference.
Using the Bekenstein--Hawking entropy $S_{\textnormal{B-H}}(M)=4\pi M^2$, one finds
\begin{equation}
\Delta S_{\textnormal{B-H}}
\equiv S_{\textnormal{B-H}}(M-\omega)-S_{\textnormal{B-H}}(M)
= -8\pi\,\omega\!\left(M-\frac{\omega}{2}\right),
\label{PW_DeltaS}
\end{equation}
and hence
\begin{equation}
\Gamma(\omega)\propto e^{\Delta S_{\textnormal{B-H}}}.
\label{PW_rate_DeltaS}
\end{equation}

\section{II. Semiclassical emission rate of the non-Hermitian lattice analogue model}
\label{contourdeform}

The classically forbidden process for our lattice model is obtained by computing the imaginary part of the action of a radially infalling light-like particle/antiparticle\cite{ParikhWilczek-PRL2000, Stalhammar-NJP2023} 
\begin{equation}
\textnormal{Im}(S^{\pm})=\textnormal{Im}\left(\int\limits_{r_{\textnormal{in}}}^{r_{\textnormal{out}}} dr \tilde{k}_r^{\pm}\right)=I^{\pm},
\label{SM1}
\end{equation}
where the momentum $\tilde{k}_r^{\pm}$ is restricted to the exceptional cone,
\begin{equation}
	\tilde{k}_r^{\pm}=\frac{\gamma f(r)}{1-f^2(r)}\pm\Biggl\{\frac{\gamma^2}{\left[1-f^2(r)\right]^2}\Biggr\}^{1/2}.
	\label{SM2}
\end{equation}
These solutions correspond to the particle ($+$) and antiparticle ($-$) channels involved in the tunneling process. We now focus on the classically forbidden processes associated with the particle channel near the horizon at $r_1$. Specifically, we consider a spontaneous radiation process in which a particle with positive frequency energy $\omega > 0$ is emitted. This implies that particles radiate outward from the black hole interior, causing the black hole mass to decrease from $M_o$ to $M_o - \omega$, which in turn leads to a shrinking of the horizon. Taking into account this change in mass, the integration limits around the horizon are set as $r_{\textnormal{in}} = r_1$ and $r_{\textnormal{out}} = r_1 - 2\omega$, where $r_1=2M_o$. Inserting $\tilde{k}^+_r$ into $\textnormal{Im}(S^{+})$, we obtain
\begin{equation}
	I^{+}=\textnormal{Im}\left[\int\limits_{r_1}^{r_1-2\omega} dr\left(\frac{\gamma f(r)}{1-f^2(r)}+\left\{\frac{\gamma^2}{[1-f^2(r)]^2}\right\}^{1/2}\right)\right].
	\label{SM3}
\end{equation}
The integral has a divergence at $r=r_1$ when $f(r_1)=\kappa_3=1$. To overcome this issue, we perform a Laurent expansion in Eq.~\eqref{SM3} around $r_1$. 
Keeping at most linear order terms in $r$ and considering that the chain is sufficiently large, i.e. $r_1\ll r_2$, we can write the imaginary part of the action for the particle channel as
\begin{equation}
	I^+=\textnormal{Im}\left(\int\limits_{r_1}^{r_1-2\omega}dr\frac{\gamma \left[1-\frac{\kappa_1-\kappa_2}{2l}(r-r_1)\right]}{\frac{1}{2r_1}\left(r-r_1\right)}
	+\left\{\frac{\gamma^2}{\left[\frac{1}{2r_1}\left(r-r_1\right)\right]^2}\right\}^{1/2}\right).
	\label{SM4}
\end{equation}
\begin{figure}[tbh]
	\centering
	\includegraphics[scale=2]{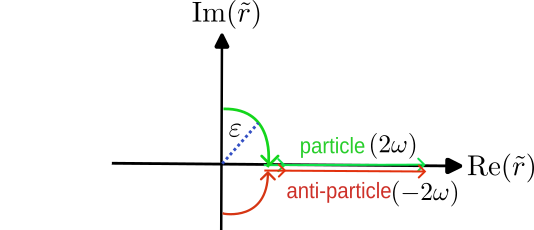}
	\caption{Contour deformation of the function $\tilde{r}$. Green (red) line represents the particle (antiparticle) channel and $\varepsilon$ represents the radius of the curve.}
	\label{contour}
\end{figure}
If we perform a change of variable $\tilde{r}=r_1-r$, Eq.~\eqref{SM4} becomes
\begin{equation}
	I^+=\textnormal{Im}\Biggl\{\int\limits_{0}^{2\omega}d\tilde{r}\Biggl[\frac{2r_1\gamma \left(1+\frac{\tilde{r}}{2r_1}\right)}{\tilde{r}}
	-\Biggl(\frac{4r_1^2\gamma^2}{\tilde{r}^2}\Biggr)^{1/2}\Biggr]\Biggr\}.
	\label{SM5}
\end{equation}
A first order pole can be found at $\tilde{r}=0$. 
The integral can be solved by deforming the contour of $\tilde{r}$ around zero, see Fig.~\ref{contour}.
Crucially, the only part of the resulting integration contour contributing to the imaginary part of the action is the (quarter) circular deformation around $\tilde{r}=0$, as the remaining line will contribute only to the real part.
By parameterizing the circular contour around $\tilde{r}=0$ as $\tilde{r}=\varepsilon e^{i\phi}$, and taking the limit $\varepsilon \rightarrow 0$, Eq.~\eqref{SM5} takes the form
\begin{equation}
	I^+=\lim_{\varepsilon\rightarrow 0}\left(\textnormal{Im}\left\{\int\limits_{\pi/2}^{0}i\varepsilon e^{i\phi}d\phi\left[\frac{2r_1\gamma \left(1+\frac{1}{2r_1}\varepsilon e^{i\phi}\right)}{\varepsilon e^{i\phi}}
	-\left(\frac{4r_1^2\gamma^2}{\varepsilon^2e^{2i\phi}}\right)^{1/2}\right]\right\}\right).
	\label{SM6}
\end{equation}
The resulting function converges uniformly on the integration contour, making it possible to move the limit inside the integration.
The integral then becomes
\begin{eqnarray}
    I^+&=& \int\limits_{\pi/2}^{0}2r_1(\gamma-|\gamma|)d\phi \nonumber \\
    &=&-\pi r_1(\gamma-|\gamma|),
    \label{SM7}
\end{eqnarray}
which leads to the solution
\begin{numcases}{I^+=}
    2 \pi r_1|\gamma| \quad &$\gamma<0$ \nonumber \\
    0 \quad &$\gamma\geq0$.
    \label{SM8}
\end{numcases}
The integral of the antiparticle channel has a similar functional form as Eq.~\eqref{SM7}, but it is important to take into account the correct integration limits [cf. the alternative deformation of the contour in Fig.~\ref{contour}, which is motivated by causality]
\begin{equation}
    I^-=\int\limits_{-\pi/2}^{0}2r_1(\gamma+|\gamma|)d\phi,
    \label{SM9}
\end{equation}
and then the solution is
\begin{numcases}{I^-=}
    2\pi r_1|\gamma| \quad &$\gamma>0$ \nonumber \\
    0 \quad &$\gamma\leq0$.
    \label{SM10}
\end{numcases}

The semiclassical emission rate is obtained by taking into account that both channels (particle and antiparticle) contribute to the probability of the tunneling process \cite{ParikhWilczek-PRL2000}. Thus, using the condition that $r_1=2M_o$, the amplitude for both channels reads  
\begin{equation}
    \mathcal{A}^{\pm}=e^{-I^{\pm}}=e^{-4\pi M|\gamma|}\Theta(\mp\gamma)
    \label{SM11}
\end{equation}
where $\Theta(x)$ is the Heaviside theta-function defined as
\begin{numcases}{\Theta(x)=}
    1 \quad &$x>0$ \nonumber \\
    0 \quad &$x\leq0$.
    \label{SM12}
\end{numcases}
Then, the semiclassical emission rate is
\begin{equation}
	\Gamma\propto (\mathcal{A}^+ + \mathcal{A}^-)^2 = (\mathcal{A}^+)^2+(\mathcal{A}^-)^2+\mathcal{A}^+\mathcal{A}^-.
    \label{SM13}
\end{equation}
The last term of the above equation vanishes because $\mathcal{A}^+$ and $\mathcal{A}^-$ do not belong to the same domain \cite{Stalhammar-NJP2023}. Therefore, the semiclassical emission rate reads
\begin{equation}
    \Gamma \propto e^{-8\pi M|\gamma|},
    \label{SM14}
\end{equation}
and the argument of the exponential can be promptly related to the Bekenstein-Hawking entropy \cite{ParikhWilczek-PRL2000, bekenstein_black_1972, bekenstein_black_1973, hawking_particle_1975}, $\Delta S_{\textnormal{B-H}}=-8\pi M|\gamma|$.

\section{III. Frequency $\omega$ in terms of the on-site potential $\gamma$}

The thermal entropy is written as
\begin{align}
    \Delta S&=\int\limits_{M_o}^{M_o-\omega}8\pi M dM \nonumber \\
    &=\int\limits_{r_1/2}^{r_1/2-\omega}4\pi\Bigl[\frac{rf^4(r)}{2}+r^2f^3(r)f'(r)\Bigr]dr.
    \label{SM15}
\end{align}

The solution of the above integral is
\begin{equation}
    \Delta S=\pi r^2f^4(r)\Bigg|_{r_1/2}^{r_1/2-\omega}.
    \label{SM16}
\end{equation}
Now, we must perform a Taylor expansion at leading order in $r$ around $r_1$. 
Then, the difference in the thermal entropy reads
\begin{equation}
    \Delta S=-\frac{3\pi}{2}\left(\frac{l}{\kappa_1-\kappa_2}\omega+\omega^2\right).
    \label{SM17}
\end{equation}
Equating the Bekenstein-Hawking entropy with Eq.~\eqref{SM17} ($\Delta S_{\textnormal{B-H}}=\Delta S$), we obtain a relation for the frequency $\omega$ of particles and antiparticles,
\begin{equation}
  -8\pi M|\gamma|=-\frac{3\pi}{2}(2M\omega+\omega^2).
  \label{SM18}
\end{equation}
In terms of the lattice parameters the frequency reads
\begin{equation}
    \omega^{\pm}=-\frac{l}{2(\kappa_1-\kappa_2)}\pm\sqrt{\frac{l^2}{4(\kappa_1-\kappa_2)^2}+\frac{8l|\gamma|}{3(\kappa_1-\kappa_2)}}.
    \label{SM19}
\end{equation}
Taking into account that $(\kappa_1-\kappa_2)/2l \ll 1$, we obtain within linear response the frequency of particles $\omega^+$ with respect to $\gamma$ by performing a Taylor expansion around zero in Eq.~\eqref{SM19}. This relation reads 
\begin{equation}
    \omega^+=\frac{8}{3}|\gamma|.
\end{equation}

\end{document}